\documentclass
[preprint,showpacs,byrevtex,10pt,twocolumn,tightenlines,prl,letterpaper]{revtex4}%
\usepackage{amsfonts}
\usepackage{amsmath}
\usepackage{amssymb}
\usepackage[dvips]{graphicx}
\usepackage{hyphenat}%
\providecommand{\U}[1]{\protect\rule{.1in}{.1in}}
\begin{document}
\preprint{ }
\title{Memory vs. irreversibility in thermal densification of amorphous glasses}
\author{Z. Ovadyahu}
\affiliation{Racah Institute of Physics, The Hebrew University, Jerusalem 91904, Israel }


\begin{abstract}
We report on dynamic effects associated with thermally-annealing amorphous
indium-oxide films. In this process the resistance of a given sample may
decrease by several orders of magnitude at room-temperatures, while its
amorphous structure is preserved. The main effect of the process is
densification - increased system density. The study includes the evolution of
the system resistivity during and after the thermal-treatment, the changes in
the conductance-noise, and accompanying changes in the optical properties. The
sample resistance is used to monitor the system dynamics during the annealing
period as well as the relaxation that ensues after its termination. These
reveal slow processes that fit well a stretched-exponential law, a behavior
that is commonly observed in structural glasses. There is an intriguing
similarity between these effects and those obtained in high-pressure
densification experiments. Both protocols exhibit the "slow spring-back"
effect, a familiar response of memory-foams. A heuristic picture based on a
modified Lennard-Jones potential for the effective interparticle interaction
is argued to qualitatively account for these densification-rarefaction
phenomena in amorphous materials whether affected by thermal-treatment or by
application of high-pressure.

\end{abstract}
\maketitle

\subsection{Introduction}

The mass-density of a solid is one of its most characteristic features. In
crystalline materials the density hardly changes even under the application of
large pressures. Yet, even a minor volume difference may induce a conspicuous
and sometime dramatic change of a measured property of the material.
Structural glasses, on the other hand, often show considerable volume change
$\Delta$V under pressure; relative volume shrinkage $\Delta$V/V exceeding 15\%
was observed in a number of studies \cite{1,2,3,4,5,6,7,8,9,10,11,12,13}.
Glasses are often amorphous structures, namely solids that lack long-range
spatial order, and it is this group of glasses that is addressed in this paper.

The increase in density under pressure was originally thought to be an
irreversible, permanent phenomenon \cite{2}. Later studies revealed hysteretic
effects in the pressure dependence of electronic \cite{6} and optical \cite{1}
properties of glasses. The observed hysteresis during a pressure cycle was
partly associated with artifacts inherent in the technique \cite{6} but there
were also indications for slow recovery from the densified phase when the
pressure was relieved in optical studies \cite{7}. These perhaps gave the
impression that a rarefied state may be the more energetically favored of the
amorphous solid.

Densification of amorphous solids by pressure is intuitively expected. It is
less trivial that volume shrinkage could also be affected by
thermal-annealing. For example, subjecting amorphous indium-oxide
(In$_{\text{x}}$O) films to a series of thermal-treatments at temperature that
exceeded the temperature T$_{\text{P}}$ at which they were prepared, resulted
in density increase by 3-20\%, depending on their composition \cite{14}. The
reduced density was reflected in the optical properties as a downward shift of
the optical-gap and concomitant increase of the refractive index while their
amorphous structure remained essentially intact \cite{14}. The relative
magnitude of these changes was comparable to the respective changes observed
in high-pressure densification experiments \cite{1,2}.

During densification, by either pressure or thermal-annealing, the
room-temperature resistivity of amorphous films may change by several orders
of magnitude \cite{6}. In the In$_{\text{x}}$O system a change in resistivity
of five-six orders of magnitude was shown to be related to the volume
shrinkage while the carrier-concentration was affected by less than a factor
of three \cite{15}. This technique has been used to study the metal-insulator
and the superconductor-insulator transitions in In$_{\text{x}}$O with various
In-O compositions \cite{16} as well as to assess the magnitude of disorder in
the insulating phase \cite{14}.

The present study started as a follow-up on the latter issue by measuring the
changes in the 1/f-noise as function of disorder that is presumably modified
during the annealing process. As will be shown in section III, the magnitude
of the flicker-noise does indeed change systematically with the resistivity of
the system. However, there appeared to be large fluctuations in the data that
led us to find, as a possible reason, a non-stationary state of the system
that follows the thermal-annealing protocol. The current work is dedicated to
the elucidation of this state an in particular to gain insight to its dynamics.

The sensitivity of the system conductivity to even a small structural change
is utilized in this work to continuously monitor the densification process
during thermal-annealing, as well as during the system relaxation that takes
place after the heating power is turned off and the sample regains its
(quasi)-equilibrium temperature. A series of \textit{in-situ} experiments
performed in this study establishes a correlation between the change in the
system resistivity and its optical transmission. This allows us to compare our
results with other glasses where only optical properties were measured, and
draw conclusions that may be pertinent to general aspects of
densification-rarefaction phenomena. A heuristic picture is presented that
offers a plausible explanation for the out of equilibrium effects that
accompany these phenomena in amorphous glasses whether driven by
thermal-treatment or application of pressure.

\subsection{Samples preparation and characterization}

The In$_{\text{x}}$O films used here were e-gun evaporated on room-temperature
substrates using 99.999\% pure In$_{\text{2}}$O$_{\text{3}}$ sputtering target
pieces. Two types of substrates were used; glass microscope-slides 1mm thick
for electrical measurements, and quartz float-glass 2 mm thick for the
\textit{in-situ} optics+resistance measurements. Deposition was carried out at
the ambience of (2-5)\textperiodcentered10$^{\text{-4}}$ Torr oxygen pressure
maintained by leaking 99.9\% pure O$_{\text{2}}$ through a needle valve into
the vacuum chamber (base pressure $\simeq$10$^{\text{-6}}$ Torr). Rates of
deposition were 0.3-0.6~\AA /s. For this range of rate-to-oxygen-pressure, the
In$_{\text{x}}$O samples had carrier-concentration \textit{N} in the range
(8-12)\textperiodcentered10$^{\text{19}}$cm$^{\text{-1}}$ measured by
Hall-Effect at room temperature. The films thickness in this study was
400-1000~\AA . Lateral sizes varied from 0.2x0.2mm$^{\text{2}}$ (for the 1/f
noise studies), 1x1mm$^{\text{2}}$ for the thermal-annealing in the small
vacuum cell, and 2x2cm$^{\text{2}}$ for the \textit{in-situ} optics-resistance
vacuum-cell. The evaporation source to substrate distance in the deposition
chamber was 45cm. This yielded films with thickness uniformity of $\pm$2\%
across a 2x2cm$^{\text{2}}$ area.

The as-deposited samples typically had sheet-resistance R$_{\square}$ of the
order of $\approx$5\textperiodcentered10$^{\text{7}}\Omega$ for a film
thickness of $\approx$10$^{\text{3}}$\AA . This was usually the starting stage
for the thermal-annealing protocols performed on each preparation batch (14
different batches were used in the study). A comprehensive description of the
annealing process and the ensuing changes in the material microstructure are
described elsewhere \cite{15,16}.

Each deposition batch included samples for optical excitation measurements,
samples for Hall-effect measurements, and samples for structural and chemical
analysis using a\ transmission electron microscope (TEM). For the latter
study, carbon-coated Cu grids were put close to the sample during its
deposition and received the same post-treatment as the samples used for
transport measurements. TEM work was only performed on two of the preparation
batches used here, mainly to check on the integrity of the procedure by
comparing with our past studies of In$_{\text{x}}$O films \cite{14,15,16}.

\subsection{Measurement techniques}

Following removal from the deposition chamber, the sample was mounted onto a
heat-stage in a small vacuum cell equipped with contacts for electrical
measurements and a thermocouple thermometer. Two cells were employed in our
study. The first (cell \#1) was used for the noise and resistance measurements
and had a light-weight heating-stage made of 0.2mm strip of copper. This
allowed for quicker changes of temperature than the one used for optics being
made of heavier 0.5mm copper sheet (cell \#2). The characteristic time to
reach 90\% of the asymptotic temperature after applying power to the
heating-stage was typically $\approx$300s for the cell \#1 and $\approx$700s
for cell \#2 used for optics measurements. Cell \#2 was installed in the
measuring compartment of the Cary-1 spectrophotometer. In addition to
electrical wires fed trough this cell for resistance and temperature
measurements it featured two quartz windows that allowed \textit{in-situ}
transmission measurements during several stages of annealing protocols. Each
optical-transmission measurement was preceded by taking a transmission scan
over the wavelength range of $\lambda$=320-860nm using a blank substrate
installed in a similar cell as that used for the actual sample. This scan was
used as a baseline to eliminate the contribution of reflections associated
with the five glass-air interfaces in the measurement set-up.

Copper wires were soldered to pressed-indium contacts to facilitate resistance
measurements. These were performed by a two-terminal technique using either
the computer-controlled HP34410A multimeter or the Keithley K617. Noise
measurements employed PAR 5113 pre-amplifier and Agilient 35670A spectrum
analyzer. The set-up involved measuring the voltage across a low-noise
resistor in series with the sample and a variable-voltage battery.

\section{Results and discussion}

\subsection{Evolution of the resistance during a thermal-annealing protocol}

The protocol we routinely use for thermal-annealing In$_{\text{x}}$O samples
is composed of the following steps: The sample, prepared at T$_{\text{P}}$
(typically$\simeq$295$\pm$2K unless otherwise specified) is anchored to a
heat-stage within the measuring cell that is then evacuated by a rotary-pump
to a pressure of $\lesssim$0.03mbar). Next the heating stage is energized, and
within a time interval $\delta$t it reaches an annealing temperature
T$_{\text{A}}$. The system is then kept at this temperature for a dwell-time
t$_{\text{d}}$, typically much longer than $\delta$t. Finally, the heat supply
is turned off and the sample is cooled back to ambient temperature within
essentially the same $\delta$t as in the heat-up stage. Figure 1 illustrates a
detailed R(t) behavior for such protocol. For brevity, the notation $\Delta
$T$\equiv$T$_{\text{A}}$-T$_{\text{P}}$ is used in all figures depicting
thermal-annealing protocols.
%TCIMACRO{\FRAME{ftbpFU}{3.039in}{2.0081in}{0pt}{\Qcb{A typical protocol used
%in thermal-annealing of In$_{\text{x}}$O films. Resistance data R(t) are shown
%in circles and refer to the left scale, the sample temperature above room's
%$\Delta$T(t)$\equiv$T$_{\text{A}}$(t)-T$_{\text{P}}$ is plotted with squares
%and refer to the right scale. The sample here has a thickness of 50nm and
%lateral dimension of 1x1mm$^{\text{2}}$. The arrows mark the onset of
%constant-temperature time-intervals starting from which fits are shown (dashed
%lines) to R(t) associated with the annealing process (8,000s\TEXTsymbol{<}%
%t\TEXTsymbol{<}86,000s) and relaxation (t\TEXTsymbol{>}92,000s). The fits to
%the data in the inset (circles for annealing, triangles for relaxation), are
%based on Eq.1 and Eq.2. These use the parameters $\beta$=0.6, 0.5 and $\tau
%$=390,000s, 8,000s respectively. To accommodate both curves in the inset, the
%data and fit for the relaxation part are shifted by a constant while the
%origin of the time scale is the position of the respective arrow.}}%
%{}{fig_1.eps}{\special{ language "Scientific Word";  type "GRAPHIC";
%width 3.039in;  height 2.0081in;  depth 0pt;  original-width 11.0177in;
%original-height 7.2428in;  cropleft "0";  croptop "1";  cropright "1";
%cropbottom "0";  filename 'Fig_1.eps';file-properties "XNPEU";}} }%
%BeginExpansion
\begin{figure}[ptb]%
\centering
\includegraphics[
height=2.0081in,
width=3.039in
]%
{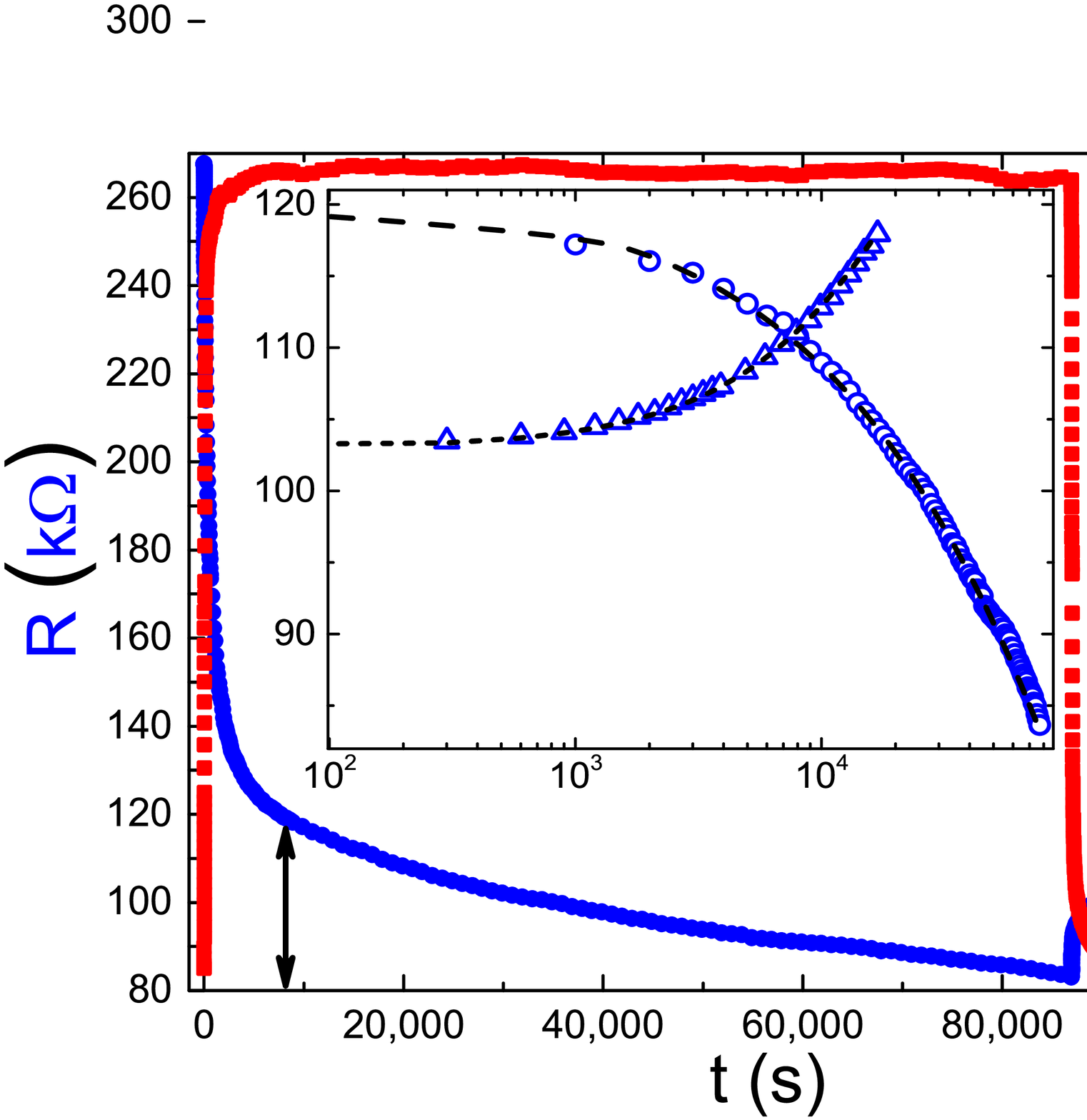}%
\caption{A typical protocol used in thermal-annealing of In$_{\text{x}}$O
films. Resistance data R(t) are shown in circles and refer to the left scale,
the sample temperature above room's $\Delta$T(t)$\equiv$T$_{\text{A}}%
$(t)-T$_{\text{P}}$ is plotted with squares and refer to the right scale. The
sample here has a thickness of 50nm and lateral dimension of 1x1mm$^{\text{2}%
}$. The arrows mark the onset of constant-temperature time-intervals starting
from which fits are shown (dashed lines) to R(t) associated with the annealing
process (8,000s$<$t$<$86,000s) and relaxation (t$>$92,000s). The fits to the
data in the inset (circles for annealing, triangles for relaxation), are based
on Eq.1 and Eq.2. These use the parameters $\beta$=0.6, 0.5 and $\tau
$=390,000s, 8,000s respectively. To accommodate both curves in the inset, the
data and fit for the relaxation part are shifted by a constant while the
origin of the time scale is the position of the respective arrow.}%
\end{figure}
%EndExpansion
120

Note first the relatively sharp response of the resistance over $\delta$t when
heating is applied and the temperature is approaching T$_{\text{A}}$, and the
similarly fast change of R during the cooling-back to T$_{\text{P}}$. These
sharp responses are mostly due to the temperature dependence of the sample
resistivity. The temperature coefficient of the resistance (TCR), for most of
the samples reported here, is negative (a sample with a \textit{positive} TCR
will be shown in Fig.6 below for comparison). The focus in this work however
is on the processes that occur while the temperature is constant, either
during the heating-period (under T$_{\text{A}}$) or after the system
temperature is back at T$_{\text{P}}$. During these time-intervals the system
resistance R evolves with time in a systematic manner:

Under T$_{\text{A}}$, the resistance decreases monotonically and $\Delta$R(T)
follows a stretched-exponential time dependence:
\begin{equation}
\Delta\text{R(t)}=\text{R}_{\text{0}}\exp\text{[-(t/}\tau\text{)}^{\beta
}\text{]}%
\end{equation}
Once the system settles back at T$_{\text{P}}$, the resistance slowly climbs
back up, also with a stretched-exponential law:%
\begin{equation}
\Delta\text{R(t)}=\delta\text{R}_{\text{a}}\text{\{1-}\exp\text{[-(t/}%
\tau\text{)}^{\beta}\text{]\}}%
\end{equation}

These functional dependences seem to be reasonably good fits to the data over
both time intervals where the sample temperature is constant if the exponent
$\beta$ is $\approx$0.5-0.6 (Fig.1). In fact, all our data for R(t) in the
relaxation regime of the annealing-cycle can be fitted to Eq.2 with $\beta$=0.5.

Stretched-exponential time dependence, with $\beta\approx$ 0.5-0.6, is
commonly observed in the dynamics of structural glasses \cite{17,18}. Several
scenarios for the origin of the stretched-exponential dependence and it
physical meaning were discussed in the literature \cite{19}. A simple
interpretation that is sometimes used suggests that the stretched-exponential
is just a weighted sum of simple exponentials reflecting an underlying
heterogeneity of the system \cite{20}; The relaxation function is the
convoluted effect of parallel relaxation events with distributed relaxation
times. It this picture the parameter $\beta$ is a logarithmic measure of the
distribution width and $\tau$ is a characteristic relaxation time \cite{20}.

Fitting data of a rather simple form such as our R(t) plots to Eq.1 or Eq.2,
involving 3 parameters is not critical enough to distinguish between possible
scenarios. These fits will be used in this work only to get an estimate for
the asymptotic value of the resistance and as rough tool to get a typical
value of the relaxation-time $\tau$ involved in the dynamics. We note in
passing that, in contrast to the electron-glass dynamics \cite{21}, these R(t)
data cannot be fitted to a logarithmic-law neither during the annealing period
nor during the relaxation that takes place after the system temperature is
reset to and stabilizes at T$_{\text{P}}$; even on half a decade in time one
observes conspicuous deviations from a log(t) dependence (see e.g., inset to
Fig.1).%
%TCIMACRO{\FRAME{ftbpFU}{3.039in}{2.1369in}{0pt}{\Qcb{Thermal-annealing
%protocol for two samples using different heating-periods. Samples are from the
%same batch with a thickness of 50nm. They were subjected to a similar
%temperature during annealing; T$_{\text{A}}$=357$\pm$1K, and T$_{\text{A}}%
%$=355$\pm$1K for (a) and (b) respectively. The arrows indicate the points at
%which T$_{\text{P}}$ has reached constant temperature.}}{}{fig_2.eps}%
%width 3.039in;  height 2.1369in;  depth 0pt;  original-width 11.0731in;
%original-height 7.7574in;  cropleft "0";  croptop "1";  cropright "1";
%cropbottom "0";  filename 'Fig_2.eps';file-properties "XNPEU";}} }%
%BeginExpansion
\begin{figure}[ptb]%
\centering
\includegraphics[
height=2.1369in,
width=3.039in
]%
{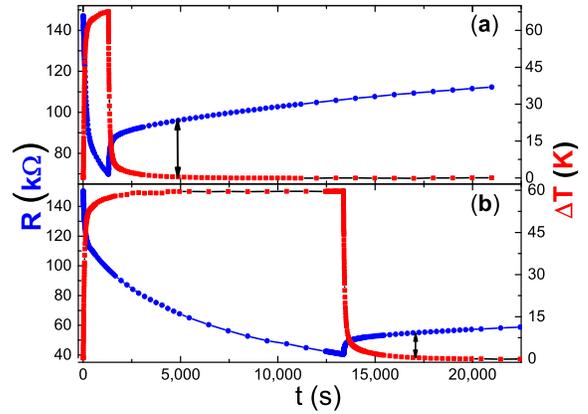}%
\caption{Thermal-annealing protocol for two samples using different
heating-periods. Samples are from the same batch with a thickness of 50nm.
They were subjected to a similar temperature during annealing; T$_{\text{A}}%
$=357$\pm$1K, and T$_{\text{A}}$=355$\pm$1K for (a) and (b) respectively. The
arrows indicate the points at which T$_{\text{P}}$ has reached constant
temperature.}%
\end{figure}
%EndExpansion

There appears to be a threshold temperature increment $\Delta$T$^{\prime}$ for
getting densification, which may be (somewhat arbitrarily) defined as the
minimum temperature increment under which a systematic reduction of the
resistance is observed during the thermal-treatment. Empirically, $\Delta
$T$^{\prime}$ depends on the specific preparation batch and on sample history.
In as-prepared samples studied here $\Delta$T$^{\prime}$ varied between
$\approx$5K to $\approx$20K. When the applied $\Delta$T was smaller than
$\Delta$T$^{\prime}$ the resistance during the heating-period may not have
changed at all even when the heating was kept for 24 hours. In one instance a
sample with R$\approx$60M$\Omega$ was kept under $\Delta$T=11K for three weeks
while R was just fluctuating around the resistance value it had at
T$_{\text{A}}$ with no sign of irreversible change.

The reduction of the sample resistance during the heating-period has been used
to change the sample resistance in studies of the metal-insulator and
superconducting-insulator transitions (MIT and SIT respectively \cite{16}). By
a judicious choice of $\Delta$T and the time of annealing one can fine-tune
the disorder and scan the immediate vicinity of the MIT (or SIT) working with
a single physical sample making this procedure an efficient technique for
these studies.

The MIT and SIT experiments naturally involve low-temperatures where the
dynamics associated with the resistance creep-up was effectively frozen. For
this reason not much attention has been given to the recovery of resistance
once the process is terminated and $\Delta$T is set back to zero. When
noticed, the slow recovery of resistance that ensued after thermal-annealing
was terminated has been ascribed to a quench-cooling effect counteracting the
annealing process. Unfortunately, it is not easy to test this conjecture
experimentally; to eliminate this effect one may need to cool the sample back
to T$_{\text{P}}$ at a \textit{much} lower rate than the typical relaxation
rate of the system, which besides being time consuming is susceptible to being
tainted by accidental artifacts.

On the other hand it can be demonstrated that, all other things being equal,
the relative amount of the resistance that is asymptotically recovered depends
in a systematic way on the time the system spends under a given T$_{\text{A}}%
$.%
%TCIMACRO{\FRAME{ftbpFU}{3.039in}{2.0781in}{0pt}{\Qcb{The ratio between the
%resistance before and after thermal-annealing as function of the dwell-time
%t$_{\text{d}}$ the sample was under a temperature of T$_{\text{A}}$=356$\pm
%$2K. Data points stand for four different samples from the same preparation
%batch (50nm thick). Sample \#1 and \#3 are the same samples shown in Fig.2 (a)
%and Fig.2 (b) respectively.}}{}{fig_3.eps}%
%width 3.039in;  height 2.0781in;  depth 0pt;  original-width 11.1699in;
%original-height 7.6043in;  cropleft "0";  croptop "1";  cropright "1";
%cropbottom "0";  filename 'Fig_3.eps';file-properties "XNPEU";}} }%
%BeginExpansion
\begin{figure}[ptb]%
\centering
\includegraphics[
height=2.0781in,
width=3.039in
]%
{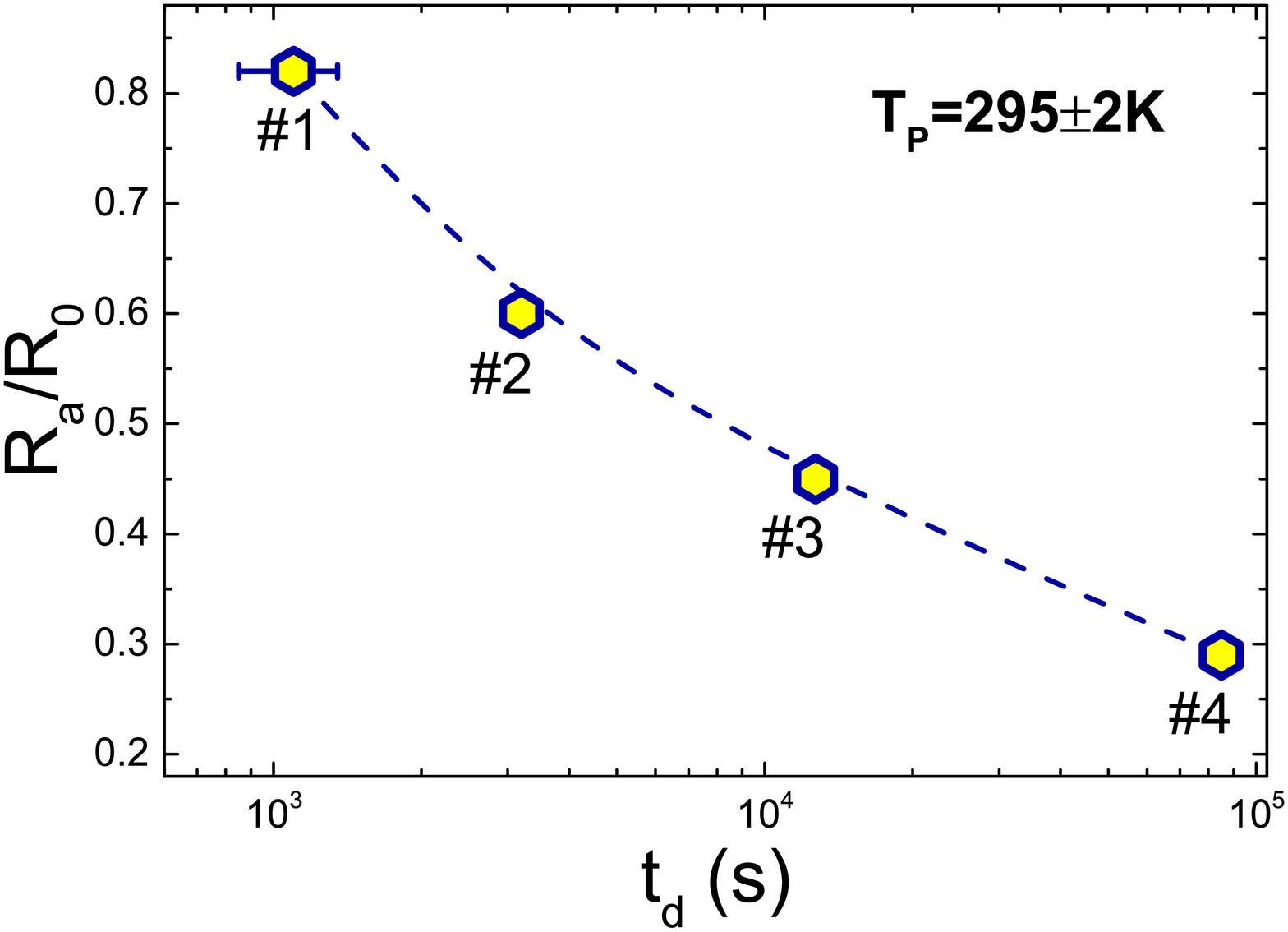}%
\caption{The ratio between the resistance before and after thermal-annealing
as function of the dwell-time t$_{\text{d}}$ the sample was under a
temperature of T$_{\text{A}}$=356$\pm$2K. Data points stand for four different
samples from the same preparation batch (50nm thick). Sample \#1 and \#3 are
the same samples shown in Fig.2 (a) and Fig.2 (b) respectively.}%
\end{figure}
%EndExpansion

An example for the role of time is shown in Fig.2: The data in the figure
illustrates that, other things being equal, shorter dwell-time results in a
larger magnitude of recovered resistance (Fig.2a). The R(T) for the final
stage of the protocol, fitted to the stretched-exponential law (yielding
similar values for $\beta$=0.5 and $\tau\approx$10$^{\text{4}}$) was used to
estimate the asymptotic values R$_{\text{a}}$=R(t$\rightarrow\infty$) for both
Fig.2a and Fig.2b data. From these one infers that thermal-annealing has
reduced the resistance by $\approx$20\% for the sample in Fig.2a and by
$\approx$120\% for the sample in Fig.2b, while the dwell-times were $\approx
$1100s and $\approx$13,000s respectively.

For a more detailed view of the role of time, one needs to quantify the change
of the resistance caused by the thermal-treatment process. A quantitative
measure of the annealing effect is the ratio R$_{\text{0}}$/R$_{\text{a}}$
between the initial resistance R$_{\text{0}}$(T$_{\text{P}}$) and the
asymptotic value for it R$_{\text{a}}$(T$_{\text{P}}$). For the latter we
extrapolate R(t) for t$\rightarrow\infty$ using Eq.2. The R$_{\text{0}}%
$/R$_{\text{a}}$ ratio as function of the annealing time is shown in Fig.3 for
several samples from the same preparation batch.

Once $\Delta$T%
%TCIMACRO{\TEXTsymbol{>}}%
%BeginExpansion
$>$%
%EndExpansion
$\Delta$T$^{\prime}$, such that some degree of densification is affected, a
finite magnitude of resistance is recovered at the asymptotic regime even when
the drop of the initial resistance during the thermal-annealing is very large
as is the case in Fig.4.
%TCIMACRO{\FRAME{ftbpFU}{3.039in}{2.1318in}{0pt}{\Qcb{Thermal-annealing
%protocol (notations as in Fig.1) for a sample with a thickness 96nm. Note the
%log scale for the R(t) data to expose the small recovered-resistance component
%that follows the termination of the heating-period. The inset shows a fit to
%the relaxation using Eq.2 with the parameters shown in the inset (origin for
%the time and for $\Delta$R are marked by the arrow).}}{}{fig_4.eps}%
%width 3.039in;  height 2.1318in;  depth 0pt;  original-width 11.4155in;
%original-height 7.9744in;  cropleft "0";  croptop "1";  cropright "1";
%cropbottom "0";  filename 'Fig_4.eps';file-properties "XNPEU";}} }%
%BeginExpansion
\begin{figure}[ptb]%
\centering
\includegraphics[
height=2.1318in,
width=3.039in
]%
{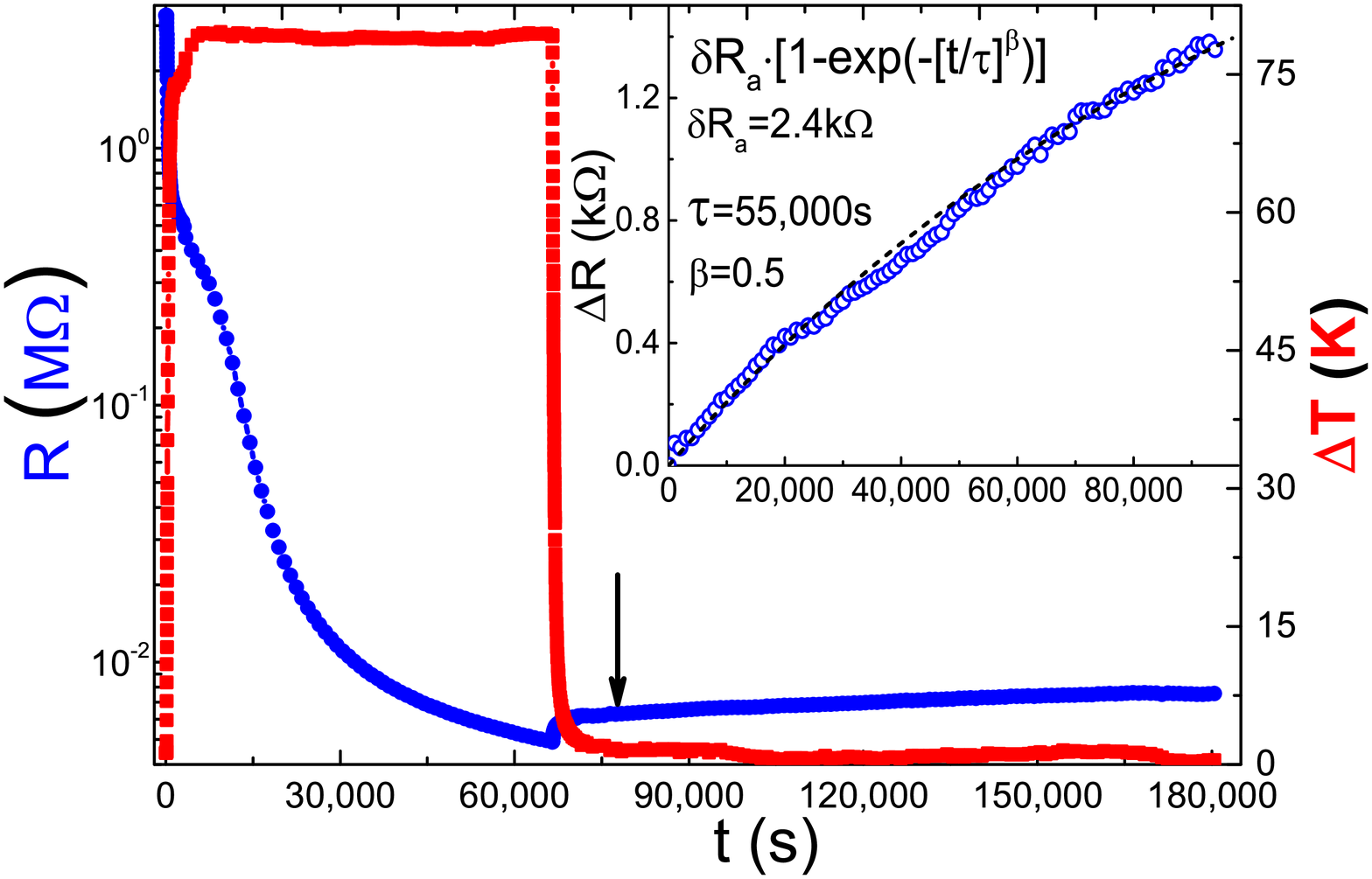}%
\caption{Thermal-annealing protocol (notations as in Fig.1) for a sample with
a thickness 96nm. Note the log scale for the R(t) data to expose the small
recovered-resistance component that follows the termination of the
heating-period. The inset shows a fit to the relaxation using Eq.2 with the
parameters shown in the inset (origin for the time and for $\Delta$R are
marked by the arrow).}%
\end{figure}
%EndExpansion
On the other extreme, a short dwelling-time, results in most of the original
resistance being recovered as depicted in Fig.5.
%TCIMACRO{\FRAME{ftbpFU}{3.039in}{2.0453in}{0pt}{\Qcb{Thermal-annealing
%protocol for a sample with thickness of 100nm. The sample has been under
%T$_{\text{A}}$ =287$\pm$2K for $\approx$250s. Its initial resistance
%R$_{\text{0}}$ was 134k$\Omega$ and its asymptotic resistance R$_{\text{a}}$
%(extrapolated through the fit to Eq.2 see inset) is 133k$\Omega$. All
%notations are as in Fig.4.}}{}{fig_5.eps}%
%width 3.039in;  height 2.0453in;  depth 0pt;  original-width 10.9347in;
%original-height 7.4633in;  cropleft "0";  croptop "1";  cropright "1";
%cropbottom "0";  filename 'Fig_5.eps';file-properties "XNPEU";}} }%
%BeginExpansion
\begin{figure}[ptb]%
\centering
\includegraphics[
height=2.0453in,
width=3.039in
]%
{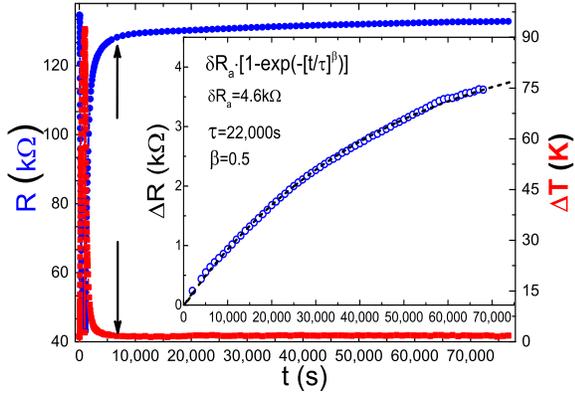}%
\caption{Thermal-annealing protocol for a sample with thickness of 100nm. The
sample has been under T$_{\text{A}}$ =287$\pm$2K for $\approx$250s. Its
initial resistance R$_{\text{0}}$ was 134k$\Omega$ and its asymptotic
resistance R$_{\text{a}}$ (extrapolated through the fit to Eq.2 see inset) is
133k$\Omega$. All notations are as in Fig.4.}%
\end{figure}
%EndExpansion

All examples for thermal-annealing shown above involve samples that are on the
insulating side of the MIT and, in particular, exhibit a negative TCR
($\partial$R/$\partial$T%
%TCIMACRO{\TEXTsymbol{<}}%
%BeginExpansion
$<$%
%EndExpansion
0) over the temperature range of our experiments. This may raise the question
of whether the slow recovery of the resistance is related to a sluggish
cooling process of the electronic system. The results of the protocol shown in
Fig.6 show that the R-recovery is independent of the TCR sign. In this case
the process of thermal annealing was extended for a long time and under a
relatively large $\Delta$T to reduce the system disorder to below the value
where the TCR changes sign and becomes positive. Still, the recovered
resistance at the end of the process has the same sign and temporal functional
dependence as when $\partial$R/$\partial$T was negative.%
%TCIMACRO{\FRAME{ftbpFU}{3.039in}{2.079in}{0pt}{\Qcb{Thermal-annealing run on a
%sample with thickness of 100nm heated to 393K for 50 hours. This is the last
%heating cycle in an annealing series that started with the sample having
%R$\approx$10M$\Omega.$ The main graph shows the last 4 hours of heating and
%the drop of R upon terminating the heating (illustrating a \QTR{it}{positive}
%TCR). The inset is an expanded view of the recovery process and shows R(t)
%monitored for almost 6 days at T=295$\pm$1K. During this process the sample
%resistance changed from $\approx$2k$\Omega$ to an asymptotic value of R=
%32.3$\Omega$.}}{}{fig_6.eps}{\special{ language "Scientific Word";
%type "GRAPHIC";  maintain-aspect-ratio TRUE;  display "USEDEF";
%valid_file "F";  width 3.039in;  height 2.079in;  depth 0pt;
%original-width 11.361in;  original-height 7.7392in;  cropleft "0";
%filename 'Fig_6.eps';file-properties "XNPEU";}} }%
%BeginExpansion
\begin{figure}[ptb]%
\centering
\includegraphics[
height=2.079in,
width=3.039in
]%
{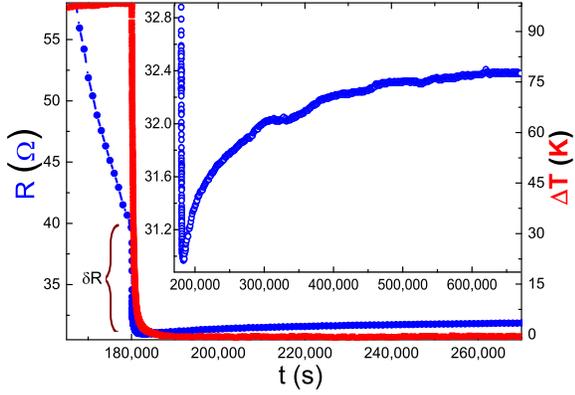}%
\caption{Thermal-annealing run on a sample with thickness of 100nm heated to
393K for 50 hours. This is the last heating cycle in an annealing series that
started with the sample having R$\approx$10M$\Omega.$ The main graph shows the
last 4 hours of heating and the drop of R upon terminating the heating
(illustrating a \textit{positive} TCR). The inset is an expanded view of the
recovery process and shows R(t) monitored for almost 6 days at T=295$\pm$1K.
During this process the sample resistance changed from $\approx$2k$\Omega$ to
an asymptotic value of R= 32.3$\Omega$.}%
\end{figure}
%EndExpansion

The behavior of the resistance during the annealing-protocol has a mechanical
analogue; The thickness of a sponge subjected to the squashing effect of a
heavy object will show qualitatively similar time-dependence as R(t) in the
examples above. In particular, it will swell back towards its original
thickness, initially as a rather fast jump followed by a much slower process.
The thickness will asymptotically recover to a degree dependent on the time
spent under the weight. In some sense, the sponge exhibits a memory of its
pristine dimensions much like the tendency of the amorphous system to recover
part of its resistance. In both cases the recovery is partial; recovery is
limited by the irreversible changes that the system incurs during
heat-treatment (or during the time the system is under pressure). These
changes are responsible for what has been referred to as "permanent"
densification \cite{1,2,3,4,5,7,8,9} and will be discussed later.

It is implicitly assumed in this work that changes in resistance occurring
while the temperature is fixed at T$_{\text{A}}$, T$_{\text{P}}$ reflect
densification, rarefaction respectively. This notion is supported by
measurements of the optical changes that take place during these processes.

\subsection{Density changes probed by optics}

The main structural change caused by thermally-annealing In$_{\text{x}}$O
films was shown to be increased mass-density by measuring the sample thickness
before-and-after the heat-treatment using x-ray interferometry \cite{14}. The
decrease of the sample volume in the process was also reflected in a downward
shift of the optical-gap and an increase of the refractive index \cite{14}.
These modified optical properties were correlated with the concomitant changes
in the system resistance. This correlation however was established for the
quasi-stationary situation, not during the time-intervals where the system is
evolving towards higher or lower density. The data shown in Fig.7 and Fig.8
were taken to test the correlation between optics and resistance during these
periods.%
%TCIMACRO{\FRAME{ftbpFU}{3.0528in}{2.0634in}{0pt}{\Qcb{A 96nm thick
%In$_{\text{x}}$O sample kept under a constant T$_{\text{A}}$=347$\pm$1K for
%5.5 hours. During this time R slowly decreased and optical transmission curves
%(inset) were taken at three different resistances at times marked by arrows
%that are coded to match the marked resistance-values.}}{}{fig_7.eps}%
%{\special{ language "Scientific Word";  type "GRAPHIC";  display "USEDEF";
%valid_file "F";  width 3.0528in;  height 2.0634in;  depth 0pt;
%original-width 11.361in;  original-height 7.6415in;  cropleft "0";
%filename 'Fig_7.eps';file-properties "XNPEU";}} }%
%BeginExpansion
\begin{figure}[ptb]%
\centering
\includegraphics[
height=2.0634in,
width=3.0528in
]%
{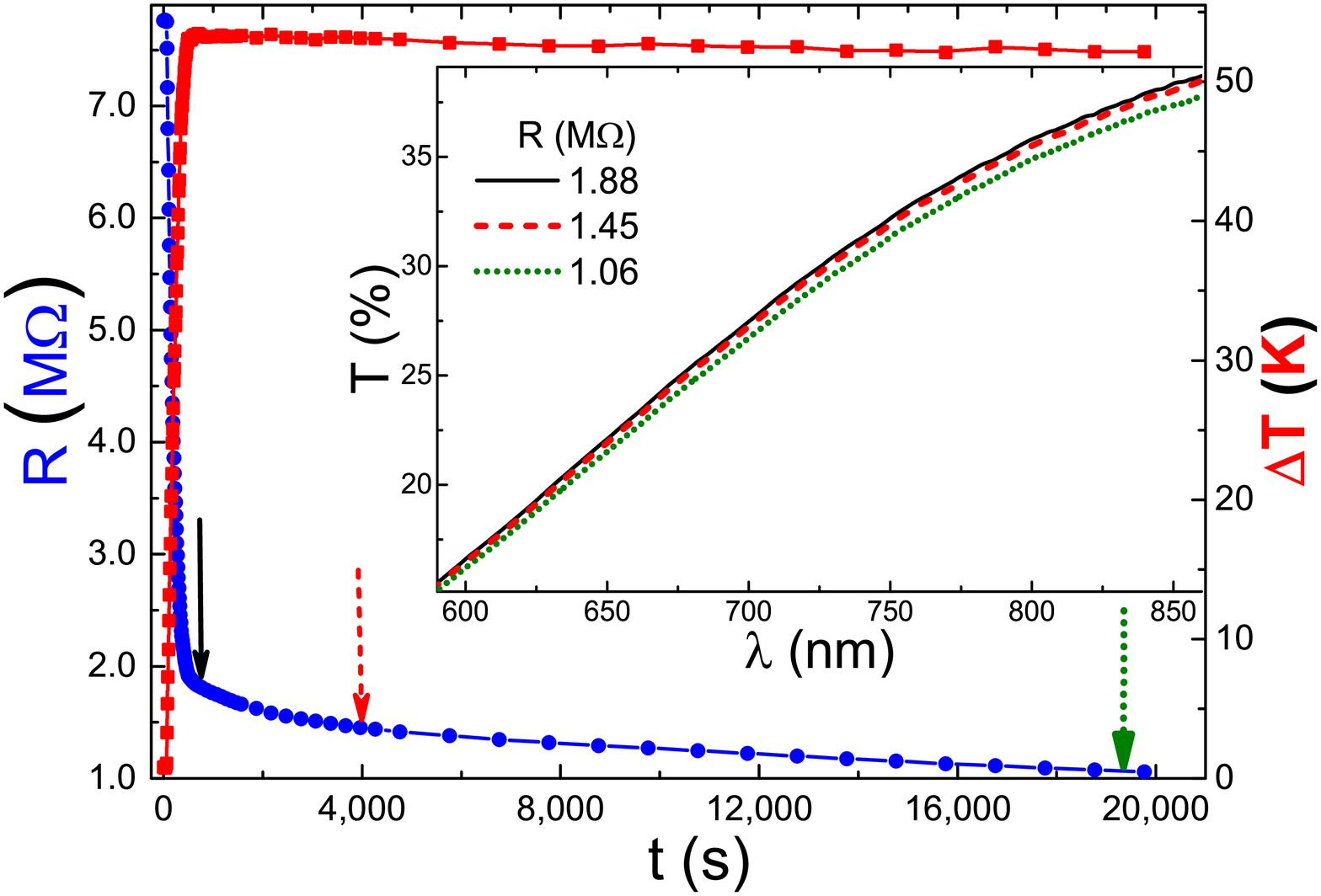}%
\caption{A 96nm thick In$_{\text{x}}$O sample kept under a constant
T$_{\text{A}}$=347$\pm$1K for 5.5 hours. During this time R slowly decreased
and optical transmission curves (inset) were taken at three different
resistances at times marked by arrows that are coded to match the marked
resistance-values.}%
\end{figure}
%EndExpansion

In these experiments, optical transmission-traces were taken at several points
in time while a constant T$_{\text{A}}$ was maintained (Fig.7) and after the
sample temperature was reset and stabilized at T$_{\text{P}}$ (Fig.8). R(t) of
the respective sample was continuously recorded during these time intervals.
Figure 8 illustrates how transmission data (Fig.9a) may be used to obtain the
shift in the optical-gap E$_{\text{g}}$ (Fig.9b) which is indicative of a
volume change; a downward shift in E$_{\text{g}}$ means densification
\cite{14}. The data in Fig.8 were taken on the same preparation batch of
In$_{\text{x}}$O as the samples in Fig.7 and Fig.8. The $\approx$85mV
reduction of the optical-gap and the $\approx$3 orders of magnitude
resistance-decrease result from the enhanced overlap of the atomic
wavefunctions during the thermal densification process \cite{14}. This
illustrates the higher sensitivity to volume change of the resistance
measurement technique. In fact, the relative change of the sample resistance
caused by thermal-annealing, is exponential with the corresponding change in
the optical transmission \cite{6}. This was confirmed here using the data for
the samples shown in Figures 7,8 and 9 (all from the same In$_{\text{x}}$O
preparation batch).
%TCIMACRO{\FRAME{ftbpFU}{3.039in}{2.0461in}{0pt}{\Qcb{A 96nm thick
%In$_{\text{x}}$O sample heated for $\approx$300s under T$_{\text{A}}$%
%=373K$\pm$1K then cooled to room-temperature and its R(t) was recorded for
%$\approx$68 hours. Optical transmission traces (shown in inset) were taken
%during the R-recovery process at times and resistance-values indicated by the
%arrows.}}{}{fig_8.eps}{\special{ language "Scientific Word";  type "GRAPHIC";
%width 3.039in;  height 2.0461in;  depth 0pt;  original-width 11.4034in;
%original-height 7.6415in;  cropleft "0";  croptop "1";  cropright "1";
%cropbottom "0";  filename 'Fig_8.eps';file-properties "XNPEU";}} }%
%BeginExpansion
\begin{figure}[ptb]%
\centering
\includegraphics[
height=2.0461in,
width=3.039in
]%
{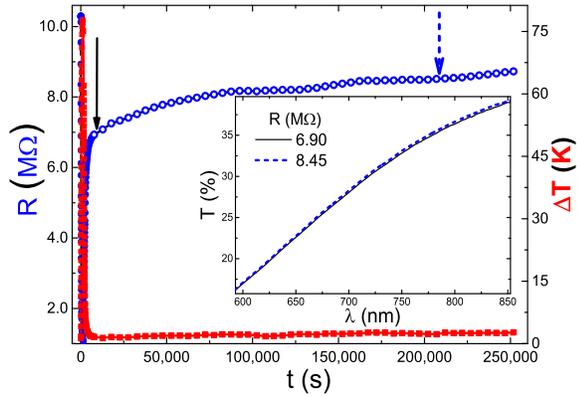}%
\caption{A 96nm thick In$_{\text{x}}$O sample heated for $\approx$300s under
T$_{\text{A}}$=373K$\pm$1K then cooled to room-temperature and its R(t) was
recorded for $\approx$68 hours. Optical transmission traces (shown in inset)
were taken during the R-recovery process at times and resistance-values
indicated by the arrows.}%
\end{figure}
%EndExpansion
The temporal resolution of the resistance measurement is also far superior to
the optics; Optical transmission measurement over the wavelength shown in
Figures 7 and 8 typically took 200s to get a well-averaged signal while the
time constant for the resistance measurement is $\approx$1s.%

%TCIMACRO{\FRAME{ftbpFU}{3.039in}{2.0133in}{0pt}{\Qcb{Optical transmission (a)
%and absorption (b) for a 96nm In$_{\text{x}}$O sample (from the same batch as
%in Figs.7\&8). Data taken at T=295K.}}{}{fig_9.eps}%
%width 3.039in;  height 2.0133in;  depth 0pt;  original-width 11.2512in;
%original-height 7.4184in;  cropleft "0";  croptop "1";  cropright "1";
%cropbottom "0";  filename 'Fig_9.eps';file-properties "XNPEU";}} }%
%BeginExpansion
\begin{figure}[ptb]%
\centering
\includegraphics[
height=2.0133in,
width=3.039in
]%
{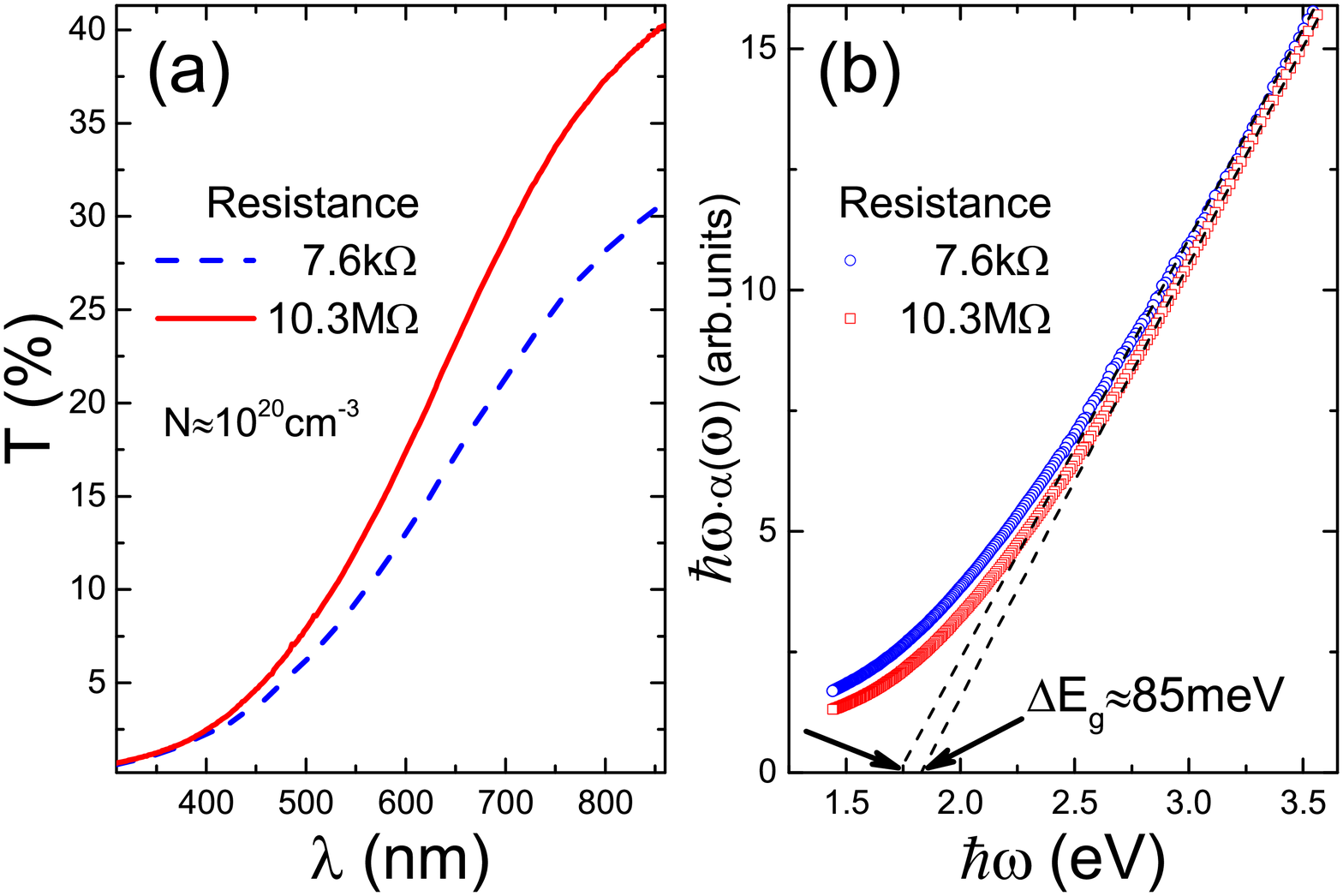}%
\caption{Optical transmission (a) and absorption (b) for a 96nm In$_{\text{x}%
}$O sample (from the same batch as in Figs.7\&8). Data taken at T=295K.}%
\end{figure}
%EndExpansion
The optics measurements in Fig.7 suggest densification, consistent with the
decrease of the sample resistance accumulated over the time the sample
temperature was fixed at T$_{\text{A}}$. Similarly, the transmission data in
Fig.8 imply that during the time the sample relaxes after its temperature
reaches T$_{\text{P}}$, the system volume and its resistance slowly increase
with time.

\subsection{Comparison with pressure-induced densification}

It is interesting to compare the effects due to thermally-annealing
In$_{\text{x}}$O films to those obtained on other glasses by applying high
pressure. We have to restrict this comparison however to the quasi-static
features of the phenomena as the details of the dynamics has yet to be
systematically studied in high-pressure experiments (presumably due to
inadequate temporal resolution inherent to the technique).

There is comprehensive literature on pressure-induced densification studies
spread over almost seven decades and involving more than ten different
materials \cite{1,2,3,4,5,6,7,8,9,10,11,12,13}. In some experiments the change
of volume was directly measured. In many studies however the property that was
monitored was taken as indicatory of the volume decrease, which was taken for
granted. In \cite{6}, for example, the resistance of amorphous As$_{\text{2}}%
$Te$_{\text{3}}$ at room temperature was this property. It was observed to
drop by 6-7 orders of magnitude with pressure of 100kBar. This is a similar
swing of resistance obtained by thermal-annealing In$_{\text{x}}$O samples, as
in the series culminating in Fig.6 above. It only takes longer to get there by
the thermal technique.

Indeed, it appears that the main difference between applying pressure and
temperature is in the associated relaxation \textit{time. }In terms of the
qualitative features of the densification-rarefaction phenomena, the
similarity between the two agents is remarkable. Let us review the main
features of the phenomena as they appear when induced by the two techniques:

\begin{itemize}
\item As alluded to above, there seems to be a threshold $\Delta$T necessary
for densification. A similar impression was formed already in early pressure
works: "\textit{A definite threshold pressure is observed in vitreous silica
and silicate glasses, under which no effect takes place and above which the
collapse takes place readily...}\emph{" }\cite{1}, to give one example.

\item There is a correlation between shrinkage of the system volume and a
downward shift of the optical-gap in thermally-annealed In$_{\text{x}}$O. The
same correlation has been recognized in several high-pressure densification
studies: \textit{"...The red shift of optical absorption edge in the visible
region that results from densification exhibited the same pressure dependence
as that observed for density..." }\cite{12}.

\item A common feature in \textit{all} densification studies, independent of
material and technique, is the asymptotic recovery of a material property that
was changed from its original value during densification. This property may be
the volume itself or, as in most cases, a property like optical-gap or
resistance that reflects the volume change. Examples for this effect are
abundant in this paper. This fact was frequently noticed in pressure studies
as well: :...\textit{when pressure is relieved... it reverts to previous
state..." \cite{6}, "...When the pressure is removed, the absorption edge
shifts back to the high-energy side, but with hysteresis. After the pressure
is removed, the absorption edge initially remains at a lower energy than the
original value, but slowly increases..." }\cite{7}\textit{,"...The densified
state is found to be metastable and decreases in density over time spans of
the order of years..."} \cite{8}.
\end{itemize}

The appearance of this rarefaction-effect in so many different glasses
suggests an generic feature, characteristic of the glassy phase. In the
following we examine a heuristic picture that may assist in understanding the
observed phenomena and in particular, explains how thermal-annealing may
induce densification as well as the reason for the difference in dynamics
between applying high-pressure and subjecting the system to an elevated temperature.

\subsection{A heuristic picture for densification-rarefaction}

The picture we consider is based on the inter-particle potential schematically
shown in Fig.10.%
%TCIMACRO{\FRAME{ftbpFU}{3.039in}{3.4705in}{0pt}{\Qcb{A schematic description
%of the effective interparticle-potential $\Phi$ versus their separation r. Two
%forms of this potential are shown as representatives from the assumed
%continuous distribution.}}{}{fig_10.eps}%
%width 3.039in;  height 3.4705in;  depth 0pt;  original-width 5.757in;
%original-height 6.5812in;  cropleft "0";  croptop "1";  cropright "1";
%cropbottom "0";  filename 'Fig_10.eps';file-properties "XNPEU";}} }%
%BeginExpansion
\begin{figure}[ptb]%
\centering
\includegraphics[
height=3.4705in,
width=3.039in
]%
{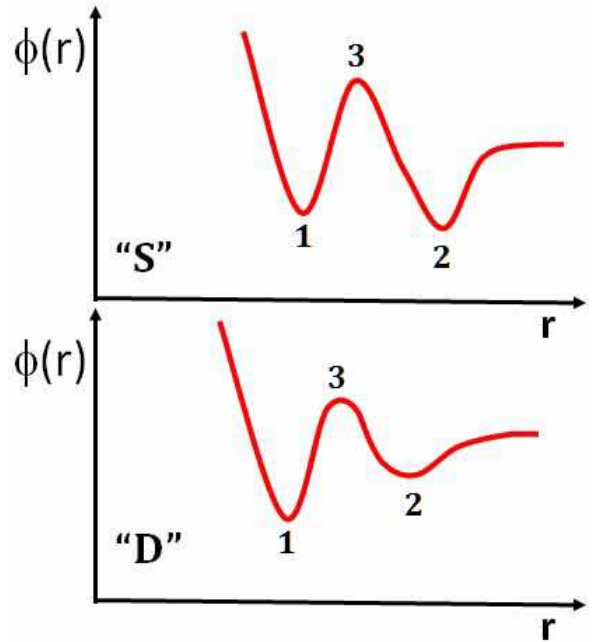}%
\caption{A schematic description of the effective interparticle-potential
$\Phi$ versus their separation r. Two forms of this potential are shown as
representatives from the assumed continuous distribution.}%
\end{figure}
%EndExpansion
The figure depicts two local configurations of the interparticle potential;
"S" and "D" are specific two-state-systems \cite{22,23} featuring two local
minima. The state labeled S (for "spongy") favors a larger interparticle
separation while D favors a dense structure. The system density at a given
temperature and pressure is determined by the $\Phi_{i}$'s. Transitions of the
type S,D(1$\rightarrow$2), S,D(2$\rightarrow$1), are assumed to be controlled
by a Boltzmann factor:%

\begin{equation}
\omega\text{\textperiodcentered exp[-}\delta\text{/k}_{\text{B}}\text{T]}%
\end{equation}
where $\delta$ is $\delta_{\text{3,1}}$=$\Phi$(3)-$\Phi$(1), $\delta
_{\text{3,2}}$=$\Phi$(3)-$\Phi$(2) respectively, $\omega\approx$%
10$^{\text{12}}$s$^{\text{-1}}$ is the attempt-frequency and T is the temperature.

Many of the local configurations in the as-made In$_{\text{x}}$O films, are
probably `spongy' because the samples were quench-condensed from the vapor
phase onto room-temperature substrates and therefore are similar to
rapidly-chilled glasses \cite{1}. Accordingly, S-configurations may initially
be preponderant in the system. When $\Delta$T%
%TCIMACRO{\TEXTsymbol{>}}%
%BeginExpansion
$>$%
%EndExpansion
0 is applied the balance of occupation in the S(1) and S(2) states changes and
the density will increase towards the level dictated by Boltzmann statistics
and controlled by the distribution of the $\delta_{\text{3,2}}$ barriers. If,
while $\Delta$T is on, {\small t}here are no irreversible structural changes
then the density will eventually saturate at the `equilibrium' value set by
the temperature. In this case the density will acquire its pristine value when
$\Delta$T is reduced to zero. Memory is preserved in this case but full
recovery of the resistance will be governed by the time the system spends
under $\Delta$T and by the $\delta_{\text{3,1}}$-barriers.

The typical relaxation time associated with the resistance recovery may be
estimated from the fitted value for $\tau$ in Eq.2. This parameter in our
R(t)-recovery data (all measured at T$_{\text{P}}$=295$\pm$2K), range between
7.7\textperiodcentered10$^{\text{3}}$s to 55\textperiodcentered10$^{\text{3}}%
$s. Using $\omega$\textperiodcentered exp[-$\delta_{\text{3,1}}$/k$_{\text{B}%
}$T] to estimate the associated typical barrier one gets $\delta_{\text{3,1}%
}\approx$1.1$\pm$0.05~eV, which is quite a reasonable value.

The situation where memory is perfect may only be realized when $\Delta$T is
very small or when applied for a very short time. Occupation of S(1)-states
for any length of time would change the interparticle interactions and some
reconstruction may locally occur. Those regions where the reconstructions
lowers the energy will sustain a modified form of the effective interparticle-potential.

Permanent densification hinges on converting S's into D's thus increasing the
weight of dense configurations in the interparticle-potential distribution.
Such a process involves reconstruction, which means that a number of atoms (of
the order of the coordination number), have to change their position. The
probability for this `many-body' process to occur while the S(1) state is
occupied, might also be represented by an effective barrier. The typical
barrier $\delta_{\text{D}}$ for densification may then be estimated using for
example the R(t) data from Fig.1; During densification R(t) was monitored for
87,000s under a constant T$_{\text{A}}$=340$\pm$2K. These data fitted to Eq.1
with $\tau\approx$3.9\textperiodcentered10$^{\text{5}}$s (inset to Fig.1)
gives $\delta_{\text{D}}$=1.5$\pm$0.1eV, somewhat larger than the typical
barrier for relaxation.

Essentially the same scenario is expected for application of pressure. The
main difference between raising the temperature or applying pressure is in the
dynamics of the\textit{ }processes; Applying pressure reduces barriers for
densification and may actually eliminate them completely \cite{11}. This would
swiftly collapse the system into the S(1) state. However, if pressure is
relieved before local reconstruction occurs, slow rarefaction should be
observed just as in the case of the $\Delta$T scenario. This slow recovery has
indeed been observed in a number of high-pressure studies \cite{8,9,13}.

Pressure is a more efficient way to achieve permanent densification as it may
be large enough to suppress $\delta_{\text{3,2}}$, and even a small shear
component (that often accompanies pressure application) is effective in
locking-in irreversible changes \cite{3}. Temperature is less effective in
this regard as there is a limit to the $\Delta$T that one may apply without
causing crystallization. For In$_{\text{x}}$O the temperature should not
exceed $\approx$450K \cite{15,24}, still much smaller than the typical barrier
involved in densification $\delta_{\text{D}}$.

Densification by thermal-annealing has many advantages over pressure,
especially for vapor deposited glasses such as the current system. It is not
an efficient method for glasses that were obtained from the melt; these
systems were subjected to T$_{\text{A}}$ extending up to the glass temperature
so they were already thermally-annealed to a certain degree. The history
associated with the cooldown from the melt would play a role in future
annealing protocols. Examples of protocols where history has a marked effect
on relaxation are illustrated in Figs.11, and 12:%
%TCIMACRO{\FRAME{ftbpFU}{3.039in}{1.9683in}{0pt}{\Qcb{A two-temperatures
%annealing protocol. The sample, 50nm thick, is heated to and maintained at
%T$_{\text{A}}$=348$\pm$2K for 1300s. Then, the temperature is changed and held
%at T$_{\text{A}}$=340$\pm$1K. The resistance versus time recorded throughout
%the protocol exhibits a non-monotonic behavior (shows on an expanded scale in
%the inset) at the later stage of the protocol.}}{}{fig_11.eps}%
%width 3.039in;  height 1.9683in;  depth 0pt;  original-width 11.361in;
%original-height 7.3215in;  cropleft "0";  croptop "1";  cropright "1";
%cropbottom "0";  filename 'Fig_11.eps';file-properties "XNPEU";}} }%
%BeginExpansion
\begin{figure}[ptb]%
\centering
\includegraphics[
height=1.9683in,
width=3.039in
]%
{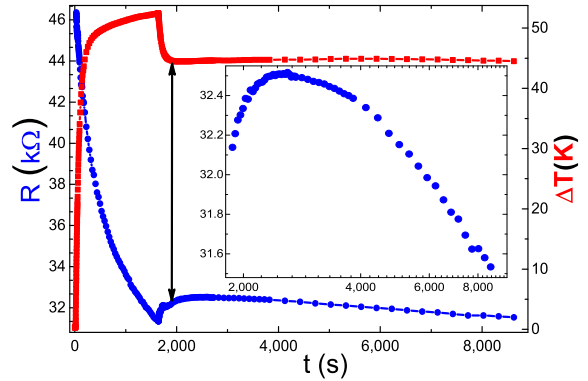}%
\caption{A two-temperatures annealing protocol. The sample, 50nm thick, is
heated to and maintained at T$_{\text{A}}$=348$\pm$2K for 1300s. Then, the
temperature is changed and held at T$_{\text{A}}$=340$\pm$1K. The resistance
versus time recorded throughout the protocol exhibits a non-monotonic behavior
(shows on an expanded scale in the inset) at the later stage of the protocol.}%
\end{figure}
%EndExpansion

These figures show R(t) for a protocol with two annealing periods at different
temperatures: after the system spends time under T$_{\text{A}}$ the
temperature is reduced and kept at intermediate temperature T$_{\text{I}}\,$,
T$_{\text{P}}$%
%TCIMACRO{\TEXTsymbol{<}}%
%BeginExpansion
$<$%
%EndExpansion
T$_{\text{I}}$%
%TCIMACRO{\TEXTsymbol{<} }%
%BeginExpansion
$<$
%EndExpansion
T$_{\text{A}}$. Such a protocol is an attempt to thermally-anneal a system at
T$_{\text{I}}$%
%TCIMACRO{\TEXTsymbol{>}}%
%BeginExpansion
$>$%
%EndExpansion
T$_{\text{P}}$ after it was thermally-annealed at T$_{\text{A}}$%
%TCIMACRO{\TEXTsymbol{>}}%
%BeginExpansion
$>$%
%EndExpansion
T$_{\text{I}}$ for a \textit{finite} time t$_{\text{d}}\lll\tau$. On the basis
of the previous experiments shown above one may expect that R(t) would exhibit
the recovery effect just after the system is cooled from T$_{\text{A}}$ and
settled at T$_{\text{I}}$. On the other hand, being above its `equilibrium'
temperature (assuming that the time spent under T$_{\text{P}}$ is much longer
than the duration of the experiment), some annealing effects should contribute
to a component of R(t) that is decreasing with time. Such a $\partial
$R/$\partial$t%
%TCIMACRO{\TEXTsymbol{<}}%
%BeginExpansion
$<$%
%EndExpansion
0 component may be appreciable enough to overwhelm the recovered resistance
effect and thus produce a non-monotonic time dependence of the sample
resistance. For the set of T$_{\text{A}}$ and T$_{\text{I}}$ used in Fig.11
such a behavior is indeed observed. In general however, the conditions under
which a non-monotonic R(t) is observable seems to require that the
intermediate temperature T$_{\text{I}}$ be closer to T$_{\text{A}}$ than to
T$_{\text{P}}$ or that the time the system spends under the influence of
T$_{\text{I}}$ is impractically long. Otherwise, the behavior of R(t) may look
like the results shown in Fig.12:%
%TCIMACRO{\FRAME{ftbpFU}{3.039in}{1.9761in}{0pt}{\Qcb{A two-temperatures
%annealing protocol. The sample, 92nm thick, is heated to T$_{\text{A}}%
%$=347$\pm$2K for 1300s and then the temperature is changed to 316$\pm$1K. The
%ensuing R(t) in this case is monotonic. Presumably the amplitude of the
%annealing component at 316K is insufficient to overcome the effect of recovery
%associated with the drop from the higher temperature. The inset shows the
%recovered resistance starting from the point (marked by the arrow in the main
%figure) where the sample temperature stabilized at T=316K. The dashed line is
%the best fit to the asymptotic region of the relaxation based on Eq.3.}}%
%{}{fig_12.eps}{\special{ language "Scientific Word";  type "GRAPHIC";
%width 3.039in;  height 1.9761in;  depth 0pt;  original-width 11.4034in;
%original-height 7.376in;  cropleft "0";  croptop "1";  cropright "1";
%cropbottom "0";  filename 'Fig_12.eps';file-properties "XNPEU";}} }%
%BeginExpansion
\begin{figure}[ptb]%
\centering
\includegraphics[
height=1.9761in,
width=3.039in
]%
{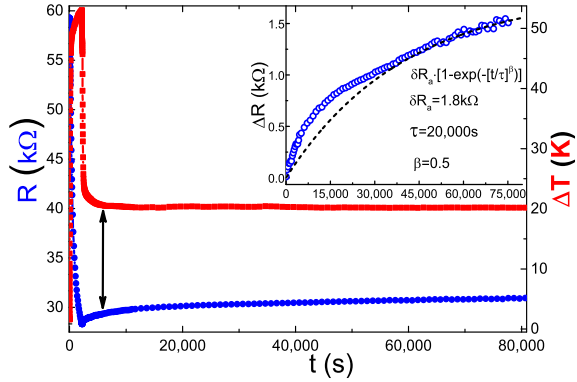}%
\caption{A two-temperatures annealing protocol. The sample, 92nm thick, is
heated to T$_{\text{A}}$=347$\pm$2K for 1300s and then the temperature is
changed to 316$\pm$1K. The ensuing R(t) in this case is monotonic. Presumably
the amplitude of the annealing component at 316K is insufficient to overcome
the effect of recovery associated with the drop from the higher temperature.
The inset shows the recovered resistance starting from the point (marked by
the arrow in the main figure) where the sample temperature stabilized at
T=316K. The dashed line is the best fit to the asymptotic region of the
relaxation based on Eq.3.}%
\end{figure}
%EndExpansion
The difference from the behavior that ensues when T$_{\text{A}}$ is switched
to T$_{\text{P}}$ as in the previous cases above (rather than to T$_{\text{I}%
}$ in Fig.12), is in the functional dependence of R(t). As the inset to Fig.12
shows, $\Delta$R(t) cannot be fitted to a single-component mechanism over the
entire time-interval where the sample was already at a constant temperature.
This is in contrast to the resistance-relaxation curves in figures 2, 4, and 5
that yielded reasonably good fit to a stretched-exponential time-dependence.

These examples make it clear that the position in time of the peak resistance
in the non-monotonous R(t) is a compounded result of the two temperatures used
in the protocol \textit{and} the time spent in each.

This behavior is quite different than that observed in the electron-glass
where a memory of time spent under a specific gate-voltage \cite{25} or under
a specific longitudinal-field \cite{26} may be simply reflected in post
relaxation. This difference is presumably related to the irreversible changes
that occur when applying $\Delta$T or pressure on structural glasses. It would
be interesting to see if something like the `simple-aging' exhibited by the
electron-glass \cite{25,26} may be observed in structural glasses if the
pressure is limited to the `elastic' regime. The logic of the heuristic
picture described above suggests that the limits of elasticity must also be
set by the time the pressure is acting on the system rather than just by its
value. Besides a test of our picture, such experiments may be useful to shed
light on the microscopics involved in the mechanical properties of amorphous solids.

\subsection{1/f noise measurements at different annealing stages}

Some aspects of the results of noise measurements presented below were the
motivation of this study. There is now however another reason to look at the
outcome of these experiments: The S-D picture described above associates
densification with modification of some two-state systems. These local
structures are often cited as the building-blocks of 1/f-noise \cite{27,28},
among other properties of disordered systems. It is then natural to find out
how flicker-noise is affected by the thermal-annealing process.

The noise experiments involved a series of consecutive annealing cycles using
three different batches of In$_{\text{x}}$O films. In each annealing stage the
sample was held at progressively increasing temperature T$_{\text{A}}$ for a
dwell-time of the order of $\approx$20 hours. T$_{\text{A}}$ was chosen such
that the resulting resistance after cooldown to room-temperature, where the
noise measurement was performed, would decrease by a factor of $\approx$2. As
a rule of thumb, for most of the resistance range, this requires tuning the
heating-power such that the sample resistance at the beginning of the
heating-period is $\approx$1/3 of its initial value. Noise measurements were
taken about one hour after the sample reached room temperature. In each
annealing stage 10$^{\text{3}}$ time-sweeps were taken to get a well-averaged
power spectrum over the frequency range 2-802 Hz. To check on time dependence,
another 10$^{\text{3}}$ time-sweeps were taken 20 minutes after the first set
to compare with the first result. Noise spectra for two resistances in an
annealing series pertaining to one of the batches studied are shown in Fig.13.%
%TCIMACRO{\FRAME{ftbpFU}{3.039in}{1.951in}{0pt}{\Qcb{The noise power-spectrum
%for two resistances in the annealing series of a 92nm thick samples. The
%dashed line depicts a 1/f law for comparison. The arrow indicates the
%frequency at which the magnitude of the noise was taken for the plot in
%Fig.14.}}{}{fig_13.eps}{\special{ language "Scientific Word";
%type "GRAPHIC";  maintain-aspect-ratio TRUE;  display "USEDEF";
%valid_file "F";  width 3.039in;  height 1.951in;  depth 0pt;
%original-width 11.3887in;  original-height 7.2774in;  cropleft "0";
%filename 'Fig_13.eps';file-properties "XNPEU";}} }%
%BeginExpansion
\begin{figure}[ptb]%
\centering
\includegraphics[
height=1.951in,
width=3.039in
]%
{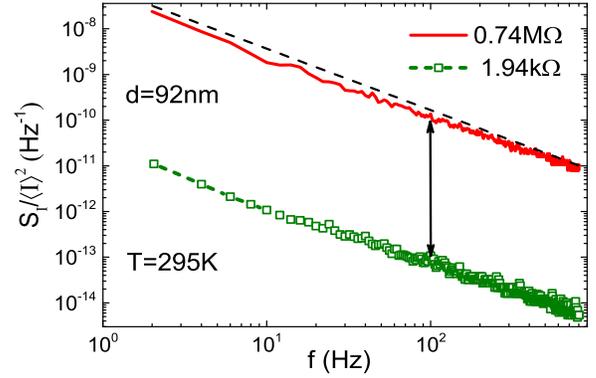}%
\caption{The noise power-spectrum for two resistances in the annealing series
of a 92nm thick samples. The dashed line depicts a 1/f law for comparison. The
arrow indicates the frequency at which the magnitude of the noise was taken
for the plot in Fig.14.}%
\end{figure}
%EndExpansion
%TCIMACRO{\FRAME{ftbpFU}{3.039in}{2.0877in}{0pt}{\Qcb{The normalized noise
%magnitude (estimated at f=100Hz, see Fig.13) as function of the sample
%resistivity. The three batches of samples that were studied in this part are
%labelled by their thickness.}}{}{fig_14.eps}%
%width 3.039in;  height 2.0877in;  depth 0pt;  original-width 10.6692in;
%original-height 7.299in;  cropleft "0";  croptop "1";  cropright "1";
%cropbottom "0";  filename 'Fig_14.eps';file-properties "XNPEU";}} }%
%BeginExpansion
\begin{figure}[ptb]%
\centering
\includegraphics[
height=2.0877in,
width=3.039in
]%
{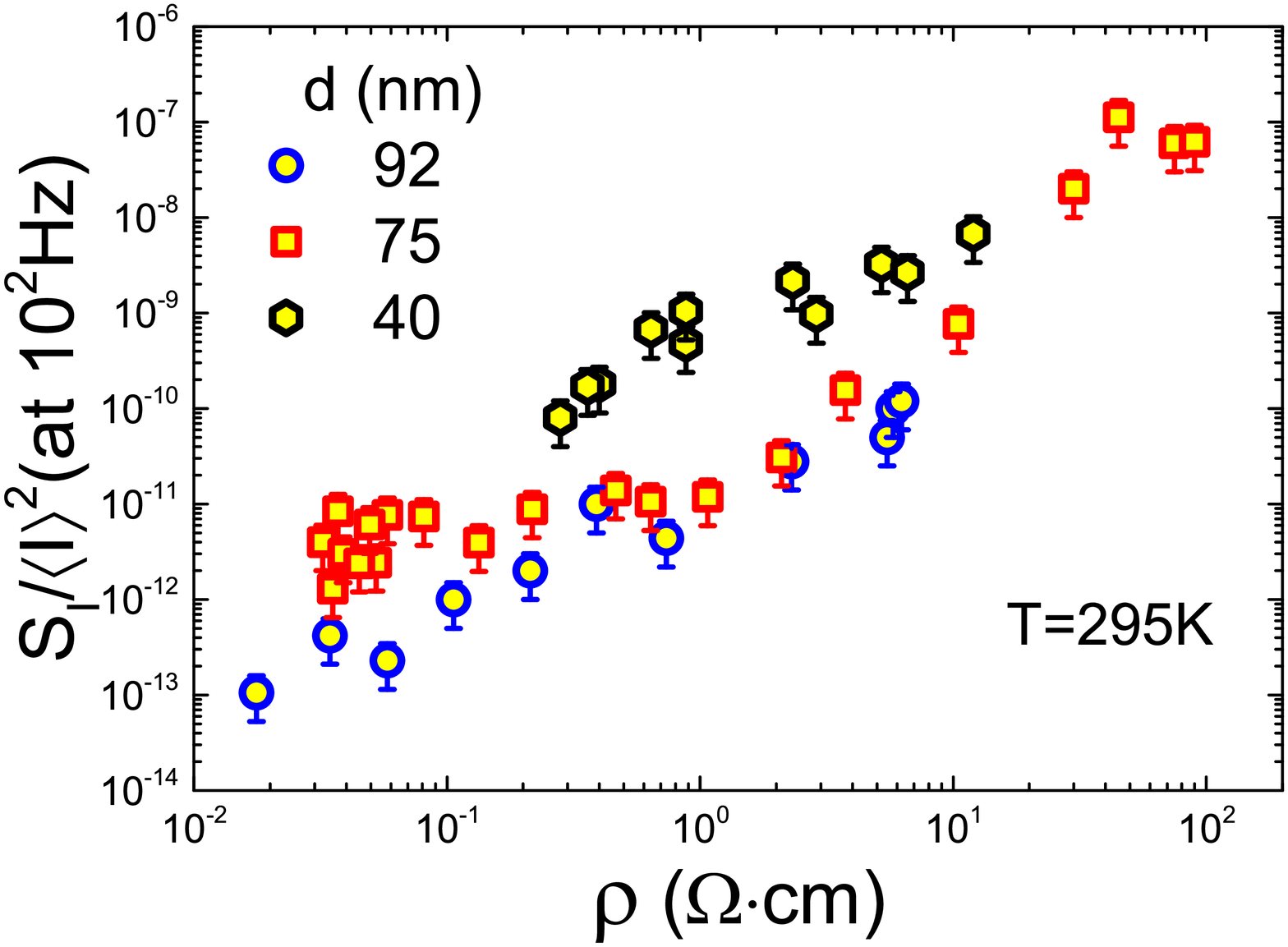}%
\caption{The normalized noise magnitude (estimated at f=100Hz, see Fig.13) as
function of the sample resistivity. The three batches of samples that were
studied in this part are labelled by their thickness.}%
\end{figure}
%EndExpansion
These traces exhibit a 1/f character except that they are slightly convex
towards the origin of the axis, a shape that characterized all our samples
independent of resistance. Data for the noise magnitude as function of
resistivity are shown in Fig.14 for the three batches of samples studied in
this work. Overall, there is a monotonic increase of S$_{\text{I}}$/$\langle
$I$^{\text{2}}\rangle$ with the sample resistivity $\rho$ but with a
considerable scatter of the data points; a factor of $\approx$5 difference in
magnitude between two consecutive measurements on the same sample was
encountered in some cases. Common artifacts relating to the measuring
circuits, non-linear effects, contacts etc. were ruled out. These large
fluctuations in the noise magnitude are apparently peculiar to
non-Gaussianity, and are often seen in 1/f studies of disordered systems
\cite{28,29,30,31}. One may suspect that the long-lasting non-equilibrium
nature of R(t) following the annealing protocol is responsible for the
deviation from a Gaussian noise \cite{28}. This issue clearly deserves further study.

The decrease on the noise magnitude with annealing may be the expected result.
Annealing is intuitively associated with removal of defects which naturally
contribute to scattering and therefore smaller resistance results in the
process. However, the experimental evidence for the structural changes only
supports a volume shrinkage that in turn leads to diminishment of the
off-diagonal disorder and broadening of the conduction band \cite{14}. The
modification of S-states to D-states in the densification process would not
necessarily influence the magnitude of the 1/f-noise either, at least not in
the frequency window probed in Fig.13. It should be noted that noise magnitude
proportional to the sample resistance has been observed in at least two other
systems: YBa$_{\text{2}}$Cu$_{\text{3}}$O$_{\text{7-}\delta}$ \cite{31} and in
some organic-composite system \cite{33}. The resistance is these systems was
changed by a variety of means; temperature, gate-voltage, oxygen-contents
(YBa$_{\text{2}}$Cu$_{\text{3}}$O$_{\text{7-}\delta}$), and by the relative
composition of the conducting compound in the organic system \cite{33}. In our
study, the system resistance was changed by yet another means - densification.
The intriguing result is that in all these cases, independent of the material
and of the way that the resistance is changed one gets the same simple
S$_{\text{I}}$/$\langle$I$^{\text{2}}\rangle\propto$R relation. As far as we
can see, the one common feature in these experiments is that the studied
systems were all in the insulating regime. It is therefore tempting to look
for the origin of this result in the general transport properties of the
hopping regime. For example, it is well known that the flicker-noise magnitude
depends on the volume of the sample \cite{27,28}. In the hopping regime, the
effective volume of the system depends on the current-carrying network which
typically occupies a small part of the physical structure and, in particular,
it depends on temperature and disorder \cite{34}. The volume associated with
this network increases as the system approaches the diffusive regime and its
resistance becomes smaller. Larger effective volume leads to a better ensemble
averaging of the contribution of local fluctuators and thus smaller
noise-amplitude. A percolation treatment of this problem, probably
incorporating correlations effects \cite{35} will be needed to find out
whether this may lead to the observed simple scaling of the noise magnitude
with the resistance. On the experimental side it would be helpful to test
other systems in order to check on the generality of the relation.

\section{Summary}

We have shown in this work that structural changes that occur during
thermal-annealing of In$_{\text{x}}$O films can be observed in real time by
monitoring the system resistance and optical properties. Using these
techniques allow tracking of the dynamics and energetics associated with these
phenomena during densification, and after its termination while the system
relaxes towards a new (metastable) state. These processes are\ activated and
exhibit temporal dependencies that fit the Kohlrausch-law often encountered in
experiments involving slow-dynamics of structural glasses.

Many of the effects produced by thermal-annealing on In$_{\text{x}}$O films
were observed in high-pressure densification experiments on a number of
amorphous systems. We have argued that these similarities may be accounted for
by assuming a simple form for the interparticle potential. A detailed study of
the dynamic associated with pressure densification should offer a stringent
test of how generic this picture may actually be. Measuring the influence of
the time spent under a constant pressure on the temporal behavior and
magnitude of the recovered volume after the pressure is released should give
valuable information on these issues.

Densification by thermal-treatment may be an effective way to shed light on
some of the fundamental properties of amorphous systems. In particular, this
technique may be used to tweak the `Boson-peak', one of the most renowned
earmarks of glasses. This feature appears as an enhanced density of
low-frequency vibrations relative to the Debye spectrum. The Boson-peak has
been conjectured to be linked to the rarefied structure \cite{36} and the
distributed nature of the interparticle separation characterizing the
disordered solid \cite{37}. More recently, it has been conjectured to be
responsible for the dominance of the Kohlrausch-law \cite{18} in glassy
dynamics. The flexibility of thermal-annealing in controlling densification
makes it a prime tool for testing these ideas.

\begin{acknowledgments}
The assistance of Dr. Netanel Aharon in data analysis is gratefully
acknowledged. This research has been supported by a grant administered by the
1030/16 grant administered by the Israel Academy for Sciences and Humanities.
\end{acknowledgments}

\end{document}